\documentclass[a4paper]{PoS}
\usepackage{bbold}
\usepackage{amsmath}

\title{Towards the heavy dense QCD phase diagram using Complex Langevin simulations}

\ShortTitle{Towards the heavy dense QCD phase diagram using Complex Langevin simulations}

\author{Gert Aarts$^1$, \speaker{Felipe Attanasio}$^{1,2}$, Benjamin J\"{a}ger$^1$, Erhard Seiler$^3$, D\'{e}nes Sexty$^{4,5}$, Ion-Olimpiu Stamatescu$^4$\\
		\llap{$^1$} Department of Physics, College of Science, Swansea University, Swansea, UK\\
		\llap{$^2$} CAPES Foundation, Ministry of Education of Brazil, Bras\'{i}lia, Brazil\\
		\llap{$^3$} Max-Planck-Institut f\"{u}r Physik (Werner-Heisenberg-Institut), M\"{u}nchen, Germany\\
		\llap{$^4$} Institut f\"{u}r Theoretische Physik, Universit\"{a}t Heidelberg, Heidelberg, Germany\\
		\llap{$^5$} Department of Physics, Bergische Universit\"{a}t Wuppertal, Wuppertal, Germany\\
        E-mail: \email{pyfelipe@swan.ac.uk}}

\abstract{Monte Carlo methods cannot probe far into the QCD phase diagram with a real chemical potential, due to the famous sign problem. Complex Langevin simulations, using adaptive step-size scaling and gauge cooling, are suited for sampling path integrals with complex weights. We report here on tests of the deconfinement transition in pure Yang-Mills SU(3) simulations and present an update on the QCD phase diagram in the limit of heavy and dense quarks.}

\FullConference{The 33rd International Symposium on Lattice Field Theory\\
		14 -18 July 2015\\
		Kobe International Conference Center, Kobe, Japan}

\begin{document}

\section{Introduction}
\begin{figure}[h!]\center
	\includegraphics[scale=0.7]{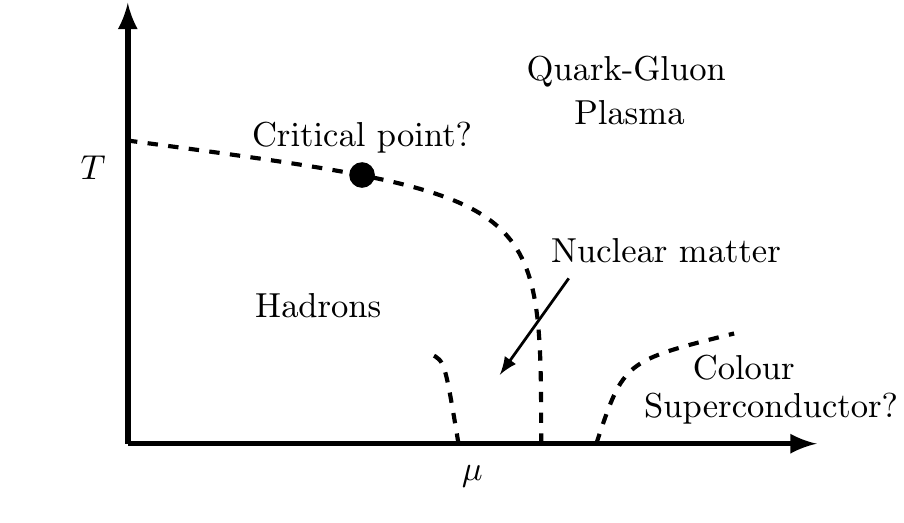}
	\vspace{-12pt}
	\caption{Sketch of the possible QCD phase diagram.}
	\label{fig.phase.diagram}
\end{figure}

The lattice formulation of QCD, in Euclidean spacetime, has enabled computer simulations of a wide variety of non-perturbative effects, such as hadron masses, decay constants and the thermal transition to the quark-gluon plasma. A full thermodynamical picture of QCD requires, however, non-zero density as well, as sketched in Figure \ref{fig.phase.diagram}. 

It is known that combining an Euclidean path integral with a non-zero chemical potential leads to the so-called \textit{sign problem}, i.e., to a complex probability weight. This leads to an exponentially hard overlap problem, which can be further worsened by numerical imprecisions. %Its oscillating part makes average values dependent on precise cancellations that can be spoiled by numerical imprecisions.

In QCD this behaviour is due to the quark determinant, after the quark fields have been integrated out
\begin{equation}
	\left[ \det M(U, \mu) \right]^* = \det M(U, -\mu^*)\,,
\end{equation}
where $U$ generically represents the gauge links and $\mu$ is the quark chemical potential.

Standard Monte-Carlo based methods can still be used in situations where the sign problem is considered ``mild'' using different techniques such as reweighting and Taylor expansion \cite{deForcrand:2010ys}, however they cannot probe reliably regions where $\mu / T \gtrsim 1$.

\section{Complex Langevin Equation}
The Complex Langevin method \cite{Klauder, Parisi:1984cs, Aarts:2009uq} is an extension of Stochastic Quantization \cite{Parisi:1980ys}. The latter consists of reproducing quantum expectation values as averages over a random walk by evolving the dynamical variables in a fictitious time $\theta$ using a Langevin equation,
\begin{equation}
	U_{x\mu}(\theta + \varepsilon) = R_{x\mu} U _{x\mu}(\theta) \,, \quad R_{x\mu} = \exp\left[ i\lambda^a(\varepsilon D^a_{x\mu}S + \sqrt{\varepsilon} \eta^a_{x\mu}) \right]\,,
\end{equation}
where $U_{x\mu}$ are the gauge links, $\lambda^a$ are the Gell-Mann matrices, $\varepsilon$ is the stepsize, $\eta^a_{x\mu}$ are white noise fields satisfying
\begin{equation}
	\langle\eta^a_{x\mu}\rangle = 0\,, \quad \langle \eta^a_{x\mu} \eta^b_{y\nu} \rangle = 2 \delta^{ab} \delta_{xy} \delta_{\mu\nu}\,,
\end{equation}
$S$ is the QCD action including the logarithm of the fermion determinant and $D^a_{x\mu}$ is defined as
\begin{equation}
	D^a_{x\mu} f(U) = \left.\frac{\partial}{\partial \alpha} f(e^{i\alpha\lambda^a} U_{x\mu})\right|_{\alpha=0}\,.
\end{equation}
In our simulations the stepsize is changed adaptively, based on the absolute value of the drift term $D^a_{x\mu} S$, in order to avoid numerical instabilities \cite{Aarts:2009dg}.
Quantum expectation values are computed as averages over the Langevin time $\theta$ after the system reaches equilibrium.

In order to deal with the sign problem we allow the gauge fields to be complex themselves. This results in enlarging the group SU($3$) to SL($3, \mathbb{C}$). In this context the gauge action is kept holomorphic by replacing $U^\dagger$ for $U^{-1}$.
With these modifications the Wilson action and the plaquette, respectively, become:
\begin{equation}
	S[U] = -\frac{\beta}{3} \sum_x \sum_{\mu < \nu} \mathbf{Tr} \left[ \frac{1}{2} \left(U_{x,\mu\nu} + U^{-1}_{x,\mu\nu}\right) - \mathbb{1} \right]\,,\quad U_{x,\mu\nu} = U_{x\mu} U_{x+\mu,\nu} U^{-1}_{x+\nu,\mu} U^{-1}_{x\nu}\,.
\end{equation}

Because of the extra degrees of freedom in SL($3, \mathbb{C}$), which is a non-compact manifold, the system might follow a trajectory where the imaginary parts of the gauge fields are no longer small deformations compared to the real ones. As a measure of how far from the unitary manifold the system is we measure the ``distance''\footnote{For U($1$) theories it is necessary to include a term of the form $(UU^\dagger)^{-1}$ to ensure $d \geq 0$. For SU($N$) this is not necessary since $\det U = 1$.}
\begin{align}
	d = \frac{1}{3\Omega} \sum_{x,\mu} \mathbf{Tr} \left[ U_{x\mu} U^\dagger_{x\mu} - \mathbb{1} \right] \geq 0\,,
\end{align}
with $\Omega$ being the lattice four-volume and equality only holding if $U_{x\mu}$ is unitary. To prevent the system from going too far from SU($3$) we use the \textit{gauge cooling} technique \cite{Seiler:2012wz}, which consists of gauge transformations constructed to push the system as close as possible to the unitary manifold in a steepest descent fashion. In other words,
\begin{equation}
	U_{x\mu} \to e^{-\varepsilon\alpha\lambda^a f^a_x} \, U_{x\mu} \, e^{\varepsilon\alpha\lambda^a f^a_{x+\mu}}\,, \quad f^a_x = 2 \mathbf{Tr} \left[ \lambda^a \left(U_{x\mu}U^\dagger_{x\mu} - U^\dagger_{x-\mu,\mu}U_{x-\mu,\mu}\right) \right]\,,
\end{equation}
where $\alpha$, similar to $\varepsilon$, is changed adaptively based on the absolute value of $f^a_x$ \cite{Aarts:2013uxa}. These transformations act orthogonally to the unitary submanifold seeking the configurations closest to it that are SL($3, \mathbb{C}$) gauge equivalent to those generated by the Langevin evolution.

Recent works include first simulation of full QCD \cite{Sexty:2013ica} and comparisons with the hopping expansion to all orders \cite{Aarts:2014bwa} and multi-parameter reweighting \cite{Fodor:2015doa}. Various discussions regarding the role of the pole of the determinant in the Complex Langevin process include \cite{Nishimura:2015pba, Splittorff:2014zca, Mollgaard:2014mga}.

\section{Tests}
In order to test the aforementioned methods we investigated the range of validity of Complex Langevin with gauge cooling, as function of the gauge coupling, and sought to reproduce the deconfinement transition in pure Yang-Mills models and compare our results to the literature \cite{Cella:1994sx}. The simulations were done in the range $0.5 \leq \beta \leq 6.5$ with an adaptive number of gauge cooling steps between each Langevin update. If the change in the distance $d$ was less than a prescribed value cooling would cease.

\begin{figure}[h!]\center
	\includegraphics[scale=0.55]{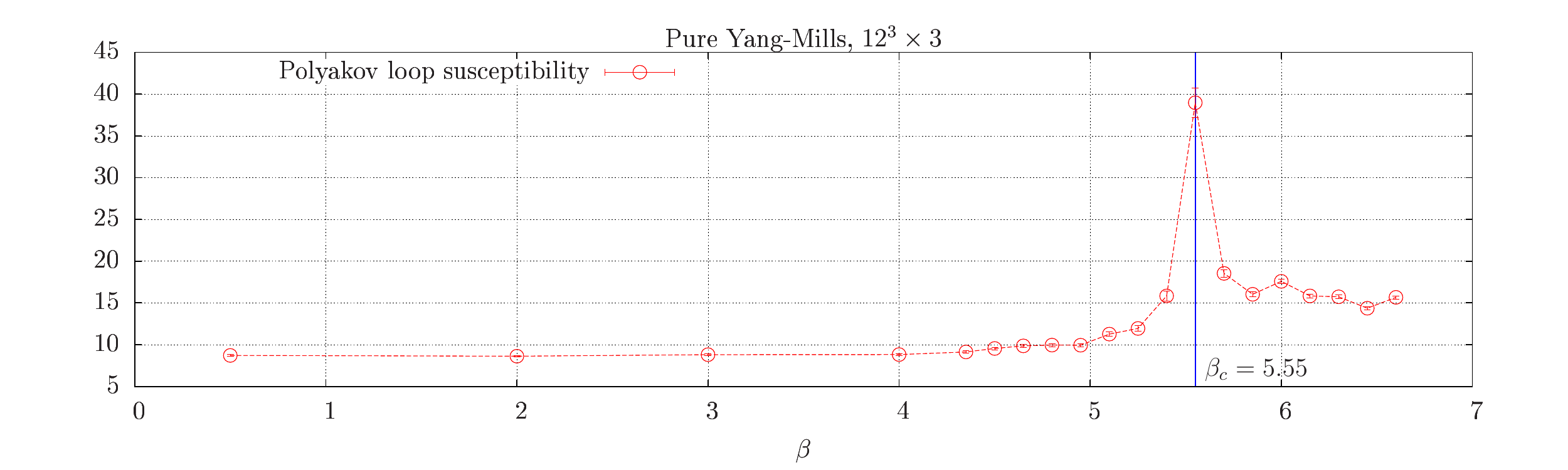}
	\includegraphics[scale=0.55]{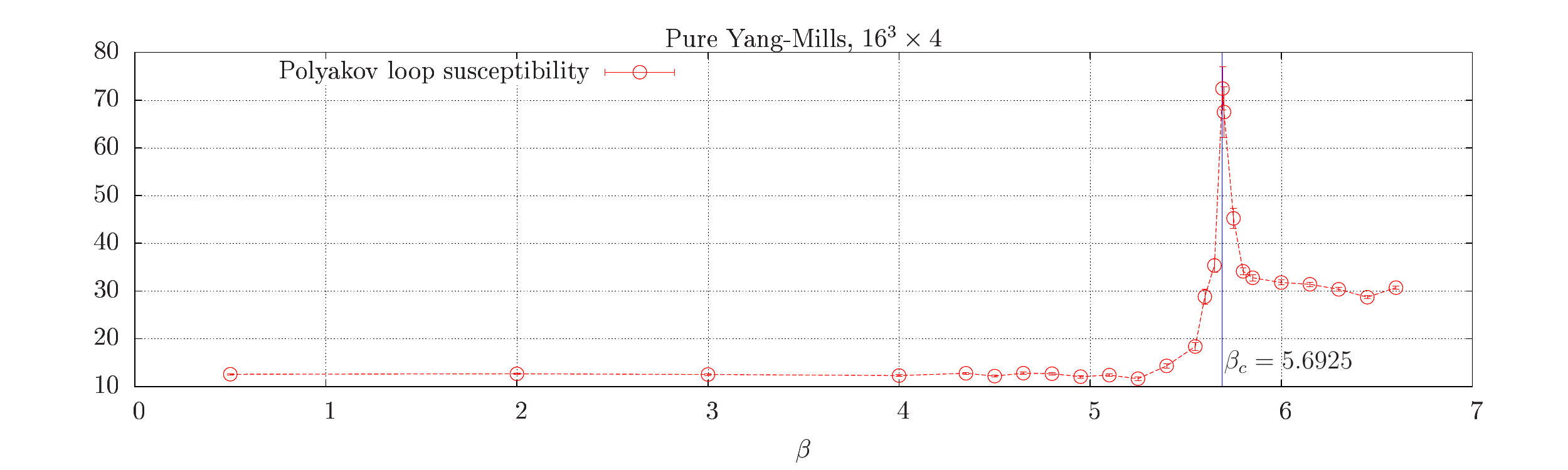}
	\vspace{-5pt}
	\caption{Polyakov loop susceptibility as a function of $\beta$ for pure Yang-Mills simulations at volumes of $12^3 \times 3$ and $16^3 \times 4$. The vertical line indicates a reference value for the deconfinement transition.}
	\label{fig.PG.polyakov.Susc}
\end{figure}

Figure \ref{fig.PG.polyakov.Susc} shows, in a wide range of $\beta$ values, the susceptibility of the traced Polyakov loop,
\vspace{-3pt}
\begin{align}
	P_{\vec{x}} = \frac{1}{3} \mathbf{Tr} \mathcal{P}_{\vec{x}} \,,\quad \mathcal{P}_{\vec{x}} = \prod^{N_\tau-1}_{\tau=0} U_4(\vec{x},\tau)\,.
\end{align}
The susceptibility peaks at the transition point, indicating deconfinement and breaking of centre symmetry.

The plots show that this range can be simulated without any problems, in agreement with expectations. The peaks of susceptibility are clearly visible and coincide with the known values indicated in the figures.

\section{Results}
The heavy-dense approximation of QCD (HDQCD) \cite{Bender:1992gn, Aarts:2008rr} is obtained by neglecting the spatial fermion hoppings, but treating the temporal part exactly:
%\vspace{-5pt}
\begin{align}
	\det M_{HD}(U,\mu) = \prod_{N_f} \prod_{\vec{x}} \left\{ \det \left[ 1 + \left(2\kappa e^\mu \right)^{N_\tau} \mathcal{P}_{\vec{x}} \right]^2 \det \left[ 1 + \left(2\kappa e^{-\mu} \right)^{N_\tau} \mathcal{P}_{\vec{x}}^{-1} \right]^2 \right\}\,,
\end{align}
where $N_f$ is the number of degenerate quark flavours and $N_\tau$ is the temporal extent of the lattice.
For a study including higher order terms in the hopping expansion, combined with a strong coupling ($\beta \to 0$) expansion, see \cite{deForcrand:2014tha}.

We have analysed the behaviour of the spatial plaquette for different gauge couplings $\beta$ as a function of the chemical potential around $\mu^2 \approx 0$: for $\mu$ purely imaginary ($\mu^2 < 0$), there is no sign problem, while for $\mu^2 > 0$ the sign problem is present. This served as a consistency check for the methods described above since the plaquette should be continuous in the vicinity of $\mu^2 = 0$.

\begin{figure}[h!] \center
	\includegraphics[scale=0.8]{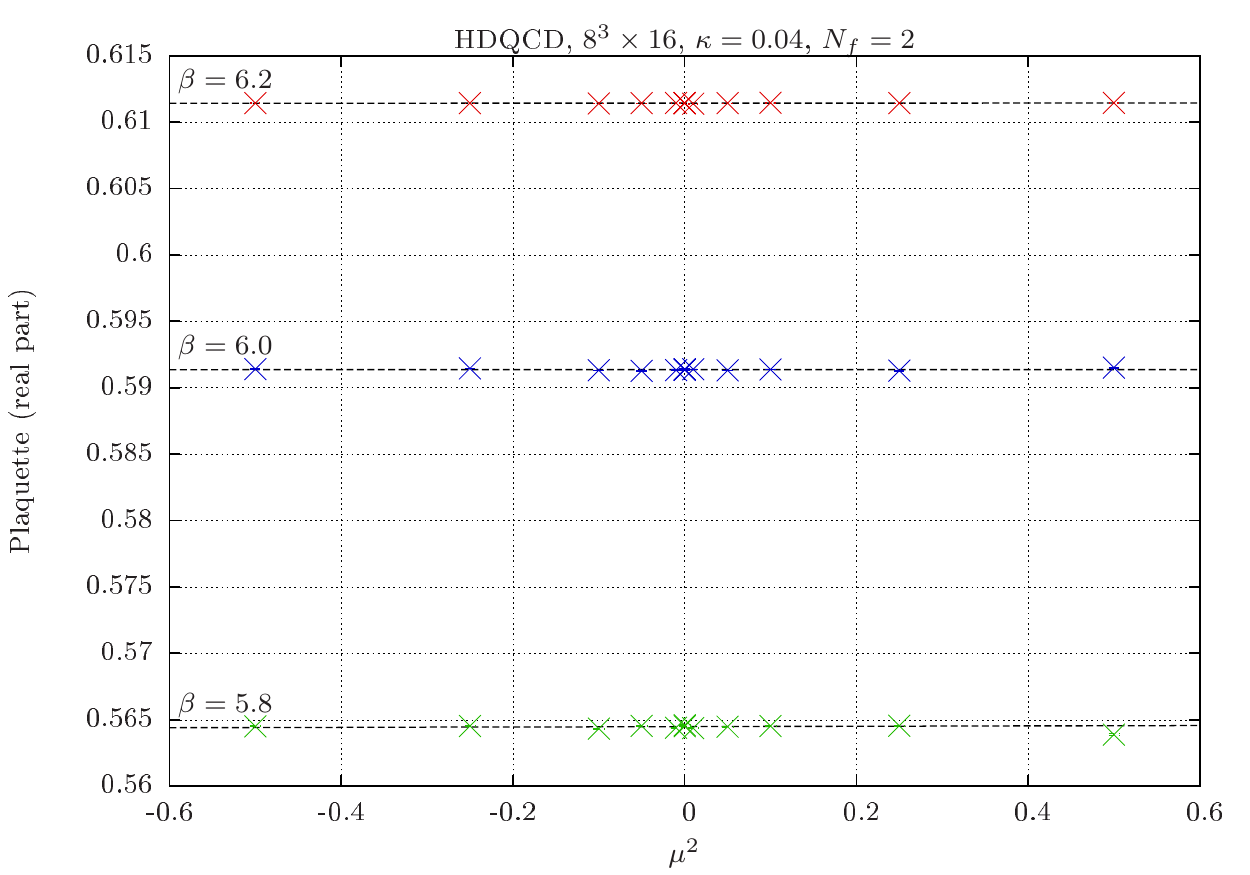}
	\vspace{-5pt}
	\caption{\label{fig.HD.allBeta}Real part of the plaquette as a function of $\mu^2$ with linear fits in the heavy-dense approximation for different gauge couplings. The $\mu^2 < 0$ region had the links periodically re-unitarised.}
\end{figure}

\begin{figure}[h!] \center
	\includegraphics[scale=0.55]{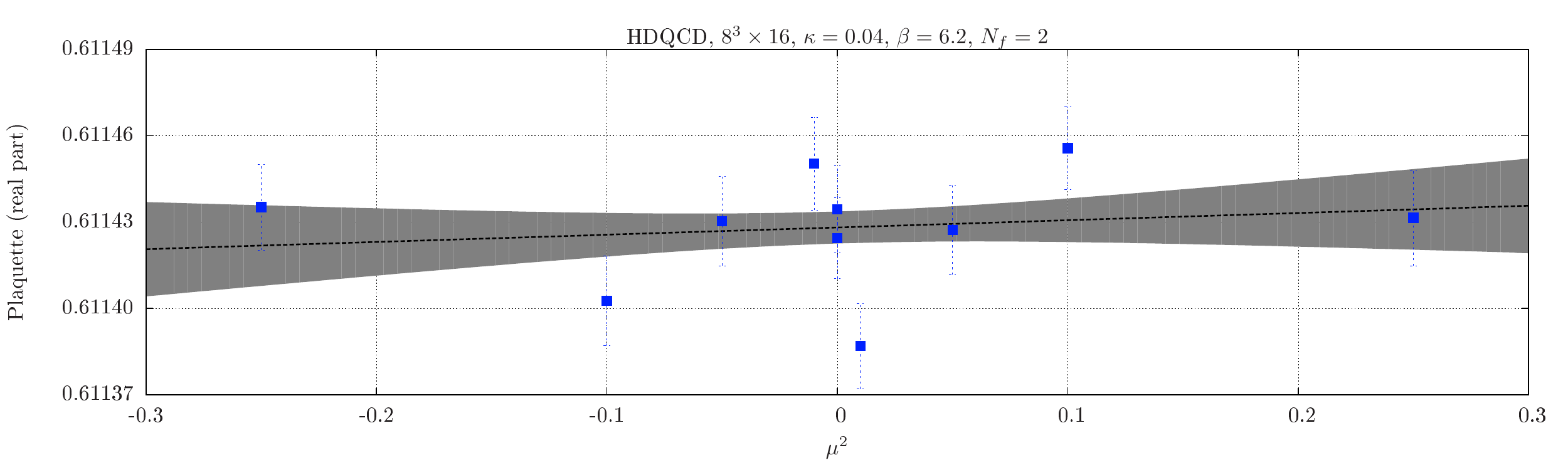}\\
	\includegraphics[scale=0.55]{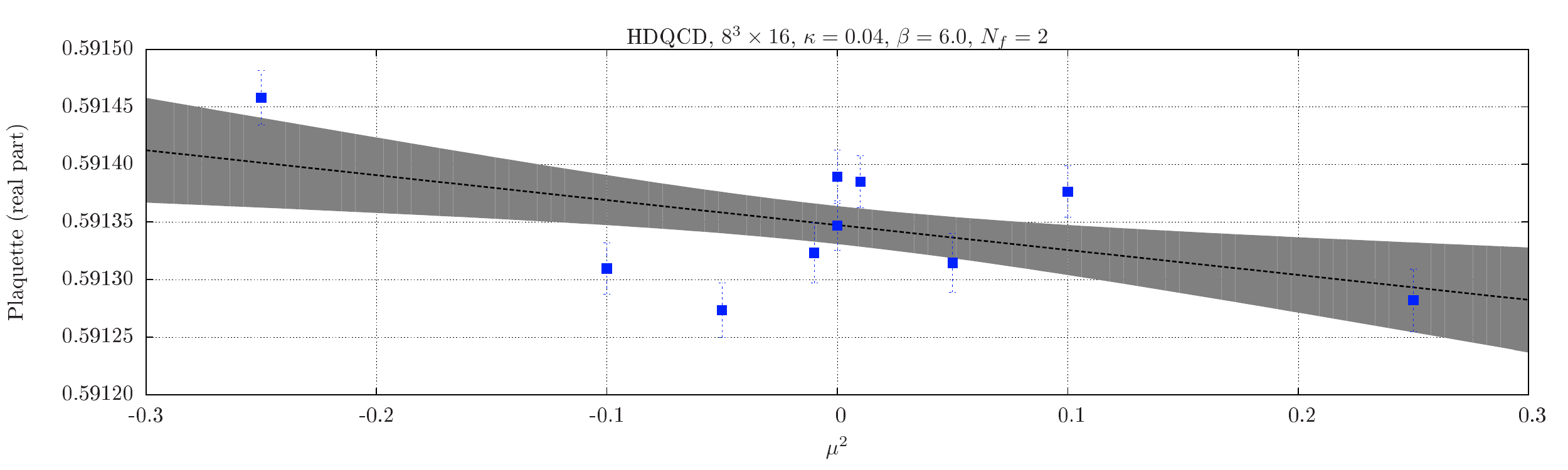}
	\vspace{-5pt}
	\caption{\label{fig.HD.zoom}Real part of the plaquette as a function of $\mu^2$ in the heavy-dense approximation for $\beta = 6.2$ and $6.0$, together with linear fits and error bands. The $\mu^2 < 0$ region had the links periodically re-unitarised.}
\end{figure}

The simulations have made use of periodical re-unitarisation of the gauge links for $\mu^2 < 0$ and gauge cooling for $\mu^2 > 0$. Our results are shown in figure \ref{fig.HD.allBeta} along with linear fits. Figure \ref{fig.HD.zoom} zooms in closer to $\mu^2 \approx 0$ for $\beta = 6.2$ and $6.0$, where it is clear that the data points are compatible with the linear fits, evidentiating the plaquette's continuity.

%\vspace{-5pt}
%\vspace{-5pt}

\begin{figure}[h!] \center
	\includegraphics[scale=0.55]{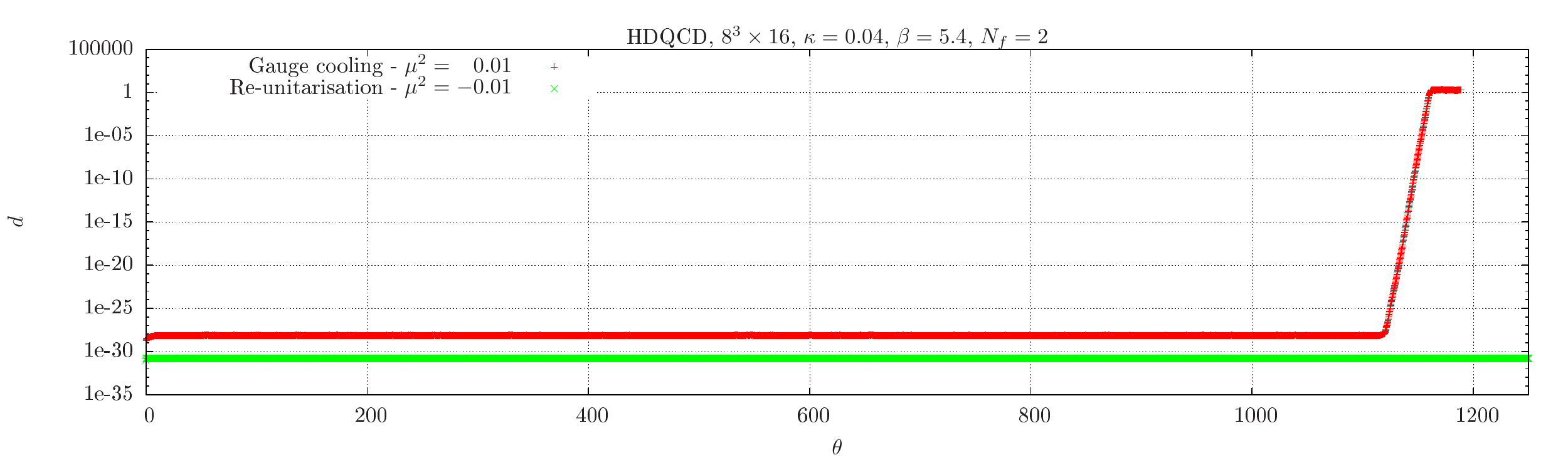}\\
	\includegraphics[scale=0.55]{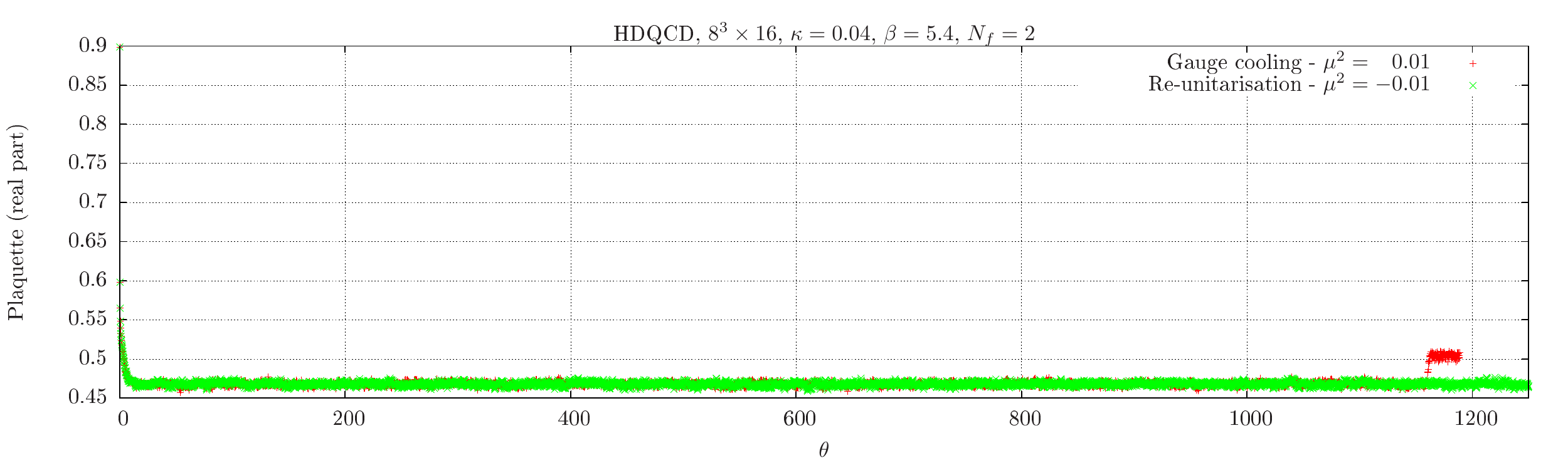}
	\vspace{-5pt}
	\caption{\label{fig.HD.norm}Distance from the unitary manifold and real part of the plaquette as functions of the Langevin time for different chemical potentials at $\beta = 5.4$.}
\end{figure}

Figure \ref{fig.HD.norm} provides an example of a situation where gauge cooling cannot prevent the distance $d$ from rising. We can see that if $d$ becomes of $\mathcal{O}(1)$ the observables exhibit a ``jump'' to another equilibrium value. It is clear that this other limit is wrong as it is not compatible with the value from re-unitarised runs in the region the plaquette should be continuous.
\vspace{-5pt}

\section{Conclusion}
We have seen that simulations using Complex Langevin equations together with gauge cooling, despite being constructed to tackle the sign problem, also work for real theories. In that situation the deconfinement transition was successfully reproduced for two different lattices without need for re-unitarisation of the gauge fields, usually employed because of numerical round-off errors.

When the sign problem is present this method avoids numerical instabilities, because of adaptive stepsize and cooling parameter, and runaway trajectories where the system would drift far into the non-compact directions of SL($3, \mathbb{C}$), due to adaptive gauge cooling. It gives consistent results around $\mu^2 \approx 0$, when transitioning from imaginary chemical potential (where no sign problem is present and the probability weight is real) to real $\mu$. In this situation the plaquette's expected continuity is verified. Care must be taken at lower $\beta$ values where gauge cooling is not able to keep the system close to the unitary manifold.

In summary, our tests show that Complex Langevin with gauge cooling is capable of dealing with pure Yang-Mills and HDQCD simulations, despite the sign problem. A full study of the phase diagram of HDQCD will be presented elsewhere \cite{Aarts:2015yba, BenLattice2015}. Dealing with potential issues that may arise around phase transitions in simpler models will certainly lead to better understanding of similar matters in the complete theory. The obvious next step is the extension to full QCD \cite{Sexty:2013ica}.

\acknowledgments
We are grateful for the computing resources made available by HPC
Wales. Part of this work used the DiRAC BlueGene/Q Shared Petaflop system at the
University of Edinburgh, operated by the Edinburgh Parallel Computing
Centre on behalf of the STFC DiRAC HPC Facility (www.dirac.ac.uk). This
equipment was funded by BIS National E-infrastructure capital grant
ST/K000411/1, STFC capital grant ST/H008845/1, and STFC DiRAC Operations
grants ST/K005804/1 and ST/K005790/1. DiRAC is part of the National
E-Infrastructure. We acknowledge the STFC grant ST/L000369/1, the
Royal Society and the Wolfson Foundation. FA is grateful for the support
through the Brazilian government programme ``Science without Borders'' under scholarship number
BEX 9463/13-5.

\end{document}